\documentclass[twocolumn,showpacs,superscriptaddress,floatfix,aps,prl]{revtex4-1}
\usepackage{amsmath}
\usepackage[dvips]{graphicx}
\usepackage{color}
\definecolor{blue}{rgb}{0,0,1.0}
\definecolor{green}{rgb}{0,1.0,0}
\definecolor{red}{rgb}{1.0,0,0}

\begin{document}

\author{N. Bachelard}
\email[Contact: ]{nicolas.bachelard@espci.fr}
\affiliation{Institut Langevin, ESPCI ParisTech CNRS UMR7587, 10 rue Vauquelin, 75005 Paris, France}

\author{J. Andreasen}
\affiliation{College of Optical Sciences, University of Arizona, Tucson, Arizona 85721, USA}

\author{S. Gigan}
\affiliation{Institut Langevin, ESPCI ParisTech CNRS UMR7587, 10 rue Vauquelin, 75005 Paris, France}

\author{P. Sebbah}
\affiliation{Institut Langevin, ESPCI ParisTech CNRS UMR7587, 10 rue Vauquelin, 75005 Paris, France}

\date{\today}

\title{Taming random lasers through active spatial control of the pump}

\begin{abstract}
Active control of the pump spatial profile is proposed to exercise control over random laser emission.
We demonstrate numerically the selection of any desired lasing mode from the emission spectrum.
An iterative optimization method is employed, first in the regime of strong scattering where modes are spatially localized
and can be easily selected using local pumping.
Remarkably, this method works efficiently even in the weakly scattering regime,
where strong spatial overlap of the modes precludes spatial selectivity.
A complex optimized pump profile is found, which selects the desired lasing mode at the expense of others, thus demonstrating the potential of pump shaping for robust and controllable singlemode operation of a random laser.
\end{abstract}

\pacs{42.55.Zz,42.25.Dd}
\maketitle

Multiple scattering of light in random media can be actively manipulated through spatial shaping of the incident wavefront \cite{vellekoopOL07}. This technique has allowed advances in focusing \cite{vellekoopNP10,cizmarNP10,badosaNP10}, and imaging \cite{popoffNC10,thompsonOL11}, paving the road to actual control of light transport in strongly scattering media \cite{vellekoopPRL08,pendryP08,popoffPRL10}.
Introducing gain in disordered media allows amplification of multiply scattered light, leading to the observation of random lasing \cite{caoLRM}. The broad range of systems where it has been studied \cite{WiersmaReview08} and the fundamental questions it has raised \cite{ZaitsevReview10,review} has captivated the community for this last decade. Prospective of random lasing is however strongly hindered by the absence of emission control: random lasers are highly multimode with unpredictable lasing frequencies and polydirectional output.
Manipulation of the underlying random structure \cite{wiersma,lawandy,wu04,souk04,wiersmaPRL,vannestepre05,lagen07,gott08,fuji09,liangAPL10}
and recent work constraining the range of lasing frequencies \cite{bardouxOE11,dardiryAPL11} have resulted in significant progress toward possible control. However, the ability to choose a specific frequency in generic random lasing systems has not yet been achieved. The spatial profile of the pump is an interesting degree of freedom readily available in random laser (e.g., \cite{leonettiNP11,kaltNP11}). In a regime of very strong scattering where the modes of the random system are spatially localized \cite{Anderson58}, local pumping allows selection of spatially non-overlapping modes \cite{VannestePRL01,sebbah02}. In weaker scattering media however (e.g., \cite{frolov99a,ling01,wu06}), several hurdles appear toward achieving fine control. Selecting modes is complicated by a narrow distribution of lasing thresholds \cite{patraPRE03,apalkovPRB05} and spatial mode overlap. Increased pumping required in these lossy systems begins to alter the random laser itself. Moreover, modifying the shape of the pump introduces changes to both the spatial and spectral properties of lasing modes
\cite{polson,wu06,wu07,andreasennu}. Such difficulties are typically absent in more conventional lasers, which have employed pump shaping,
both electrically \cite{gmachl98,hent09,shinoharaPRL10,stoneN10} and optically \cite{naidooOC11}, to select favorable lasing modes.
The question is, can shaping of the incident pump field achieve taming of random lasers?

In this letter, we exercise control over the distribution of lasing thresholds via the pump geometry to choose the random laser emission frequency. The fluctuation of lasing thresholds from mode to mode was recently found to increase when pumping a smaller spatial region \cite{andreasenpp}. Here we demonstrate numerically that the threshold of any single lasing mode for the structures investigated can be significantly separated from its neighbors if the spatial profile of the pump is correctly chosen. 
An iterative approach is proposed, inspired by spatial shaping methods employed for coherent light control \cite{vellekoopOL07} in complex media. The optimization algorithm is based on a simple minimization criterion and can be easily implemented in experiments. We first show that the algorithm simply converges to the expected localized pump profile in the localization regime. In the weakly scattering regime, we show that mode selection is also possible despite the strong mode overlap and allows for effective monomode laser action. Strikingly, we find that the evolution of the pump profile during the optimization process alters the random laser in an advantageous way, so as to achieve the desired control.

\begin{figure}[ht!]
\begin{center}
  \includegraphics[scale=0.55]{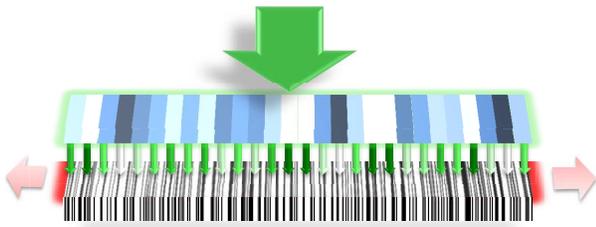}
  \caption{\label{Fig1}
    (Color online) Principle of active control of a 1D random laser.
    The black slabs represent the dielectric material which is also the gain medium.
    Optical pumping (green arrows) is transverse to the structure and amplitude modulated.
    This spatial modulation may be provided in a real experiment by a spatial light modulator (blue).
    Lasing (red arrows) occurs along the structure and will depend on the pump profile.
  }
\end{center}
\end{figure}

A one dimensional random laser is represented by a stack of 161 dielectric layers (optical index $n_1$)
separated by air gaps ($n_0=1.0$) (see Fig.~\ref{Fig1}).
Randomness is introduced in the thickness of each layer $d_{0,1} = \langle d_{0,1}\rangle (1+ \eta \zeta)$, where $\langle d_{0,1}\rangle $ is the average thickness, $0< \eta<1$ the degree of randomness, and $\zeta \in [-1;1]$ is a random number.
The position along the system is $x \in[0;L]$, where $L$ is the total length.
In the following, $\langle d_1\rangle  = 100$ nm for the dielectric layers and $\langle d_0\rangle  = 200$ nm for air,
giving a total average length $\langle L\rangle  = 48.1$ $\mu$m.
The degree of randomness is set to $\eta = 0.9$.

We choose the dielectric material as host to the gain medium described by a frequency dependent susceptibility \cite{andreasennu}
\begin{equation}
  \chi_g (k) = \frac{A_e N_{pump}}{k_a ^2 - k ^2 -i k \Delta k_a},
\end{equation}
where $A_e$ is a material-dependent constant, $N_{pump}$ is the density of excited atoms when the system is uniformly pumped,
$k_a$ is the atomic transition and $\Delta k_a$ is the spectral linewidth of the atomic resonance.
As a result, the refractive index of the dielectric becomes complex and frequency dependent, $n_1 \rightarrow \tilde{n_1}(k) = \sqrt{n_1 ^2 + \chi_g (k)}$.
In the following, the transition frequency is $k_a = 10.25 $ $\mu$m$^{-1}$ and the spectral linewidth $\Delta k_a = 0.25$ $\mu$m$^{-1}$.
The gain is assumed to be linear in the sense that it does not depend on the electrical field intensity, an assumption that is only valid at or below threshold.
We use the transfer matrix method \cite{SoukoulisPRE02,andreasennu} to find the lasing frequency, the threshold and the spatial distribution of the lasing modes.

When partial pumping is employed, the density of excited atoms due to the pumping process is modulated according to the function $f_E$ called \emph{the pump profile} as sketched in Fig.~\ref{Fig1}.
It can be written $f_E(x)\times N_{pump}$, where $x$ is the position of the layer.
The function $f_E(x)$ fulfills the constraint $0 \leq f_E(x) \leq 1$ to mimic, for instance, the amplitude modulation of the pump beam by a spatial light modulator.
This pump profile changes the gain provided by each dielectric layer, giving possible control over the lasing modes of the random laser.
Here, we aim at selecting a particular lasing mode by optimizing the pump profile.
Experimentally, a lasing mode $i$ will be selectively excited if its threshold, $N_i$, is sufficiently low and significantly lower than that of all other modes.
Hence, we introduce the rejection rate, $RR_i = \frac{min_{j \ne i} (N_{j})}{N_i}$, which compares the threshold of mode $i$ with that of the mode with the lowest threshold, $min_{j \ne i} (N_{j})$.
Selection of mode $i$ is achieved when $RR_i > 1$ provided its threshold, $N_i$ remains reasonably low.
We therefore minimize $1/RR_i+\alpha N_i$ , with $\alpha$ properly chosen to balance each term, and apply iteratively the following algorithm.
A new pump profile $f_E(x)$ results in new lasing modes and new values of $RR_i$ and $N_i$.
We apply the projected gradient method to $f_E(x) \rightarrow 1/RR_i+\alpha N_i$.
At each iteration, the gradient of $1/RR_i+\alpha N_i$ is computed.
Then, $f_E(x)$ is tuned within $[0,1]$ in the direction where the projected gradient of $1/RR_i+\alpha N_i$ is minimal.
Convergence is assumed if its relative variation is less than $10^{-4}$.

Depending on the value of the index contrast $\Delta n = n_1 - n_0$, the random laser is studied either in strongly scattering regime,
where light is confined well within the random system, or in the weakly scattering regime, where modes are extended.
For an index contrast $\Delta n = 0.60$, we find a localization length $\xi \approx 1.7 $ $\mu$m $\ll L$
over the spectral range $10.0$ $\mu$m$^{-1} < k < 10.5$ $\mu$m$^{-1}$.
The system is therefore in the localized regime and lasing modes are spatially confined within the system.
As a rule, mode selection is rather easy to achieve in this case \cite{VannestePRL01,sebbah02} using local pumping, from the initial knowledge of the location of the quasimodes of the passive system.
The localized case serves here as a test case for our iterative algorithm to check if modes can be selected without any prior knowledge of their spatial location.

\begin{figure}[ht!]
		        \centering
      			 \includegraphics[width=0.8\linewidth]{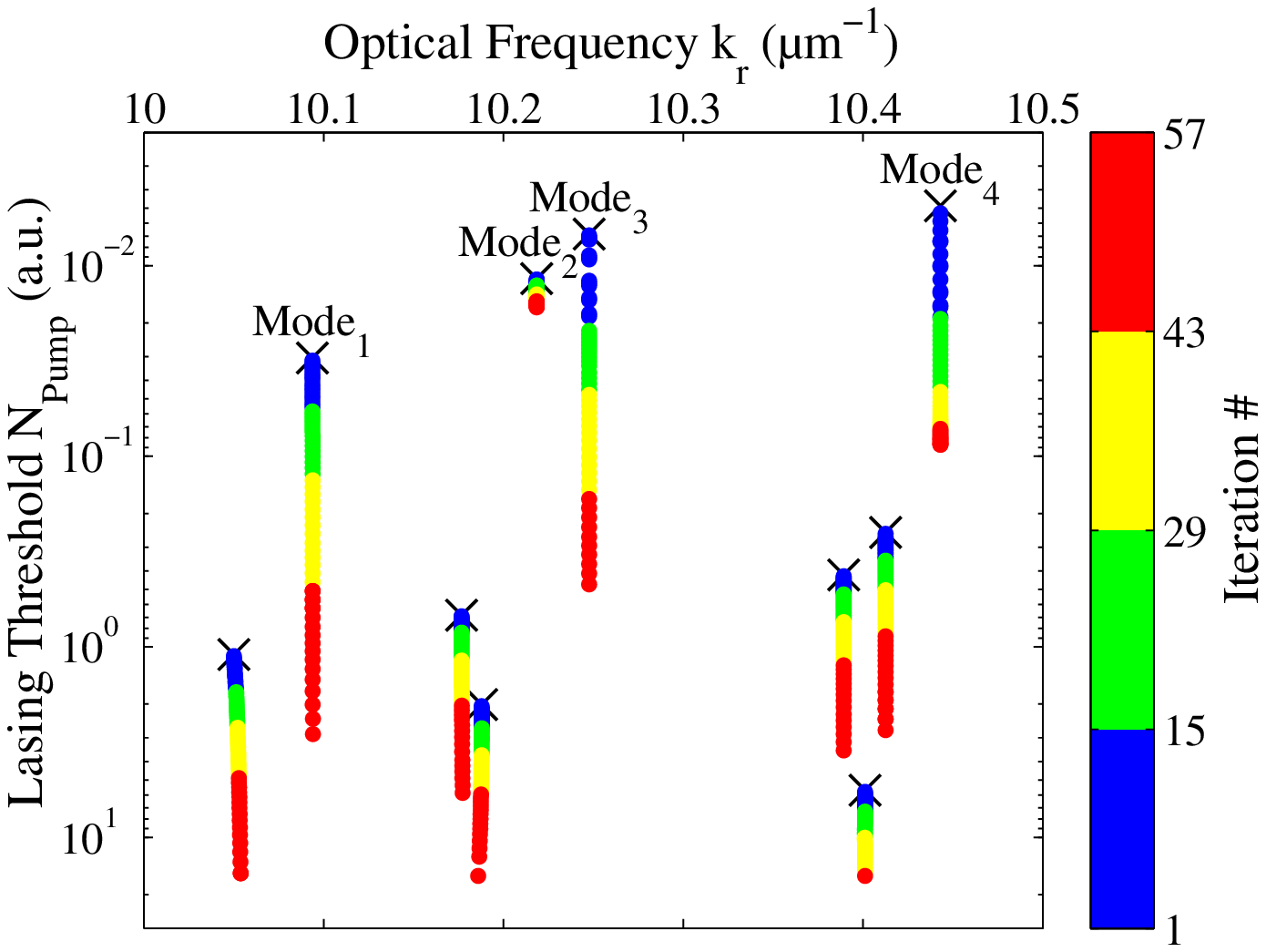}
			\centering
      			\includegraphics[width=0.8\linewidth]{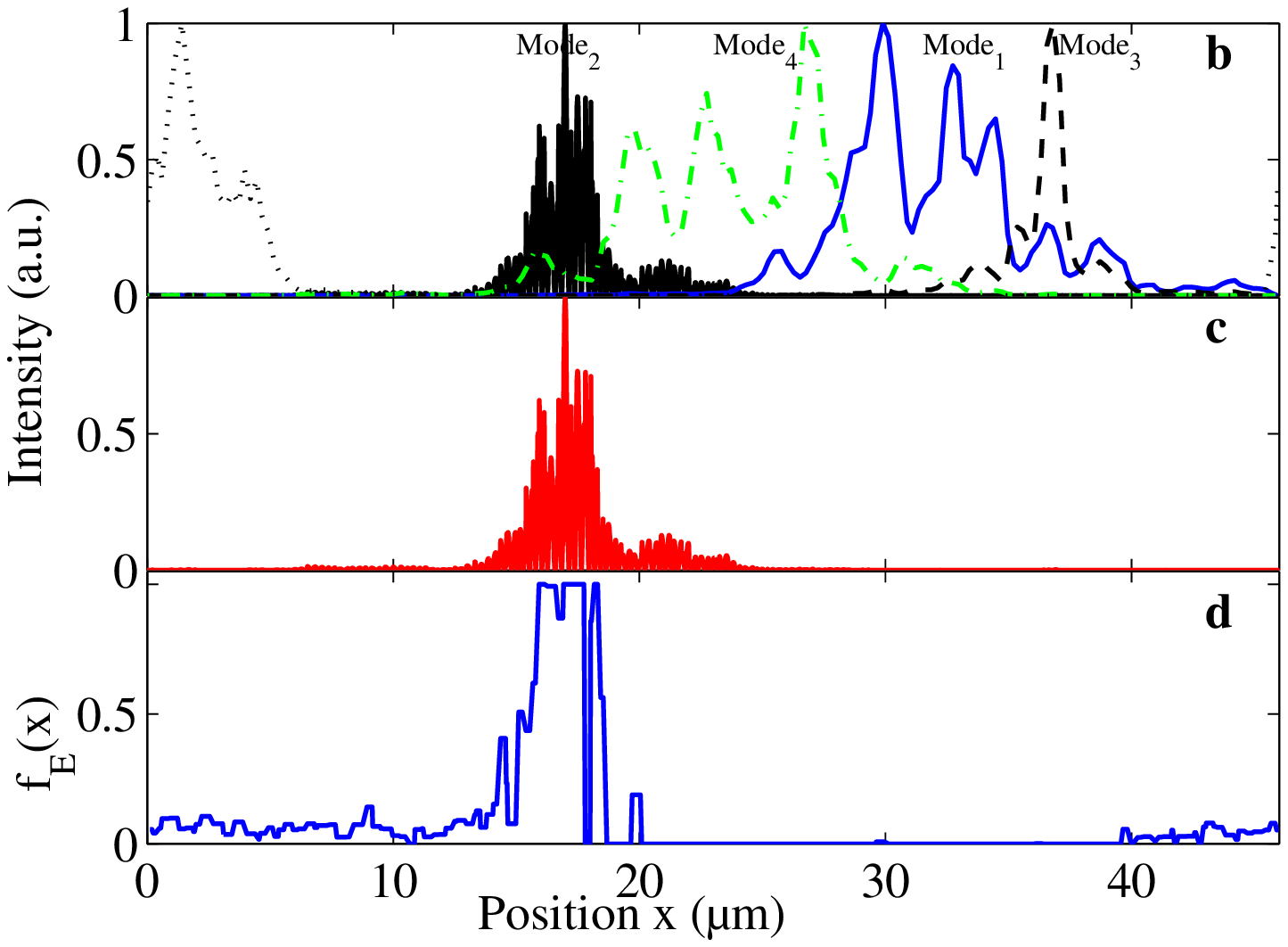}
       \caption{\label{Fig2}
         (Color online) Localized random laser case.
         (a) Frequency vs. threshold of the lasing modes at successive iterations, when optimization routine is applied to select $\text{Mode}_{2}$.
         Crosses represent initial positions for uniform pumping.
         (b) Spatial profile of five localized lasing modes. The mode profile on the left edge corresponds to the leaky mode with the highest threshold at $k = 10.4$ $\mu m^{-1}$.
         (c) Spatial profile of $\text{Mode}_2$ after optimization.
         (d) Resulting optimized pump spatial profile.
       }
\end{figure}

We first compute the threshold and optical frequency of the lasing modes for a uniform gain profile, $f_E(x) = 1$.
They are positioned in the frequency-threshold plane as crosses in Fig.~\ref{Fig2}a.
Four lasing modes with reasonably low thresholds and partial spatial overlap are chosen for demonstration.
Their spatial profiles are shown in Fig.~\ref{Fig2}b, together with the profile of a high threshold lasing mode at $k = 10.4$ $\mu$m$^{-1}$, associated with a lossy mode strongly coupled to the left end of the sample.

\begin{figure}[ht!]
\begin{center}
  \includegraphics[scale=0.55]{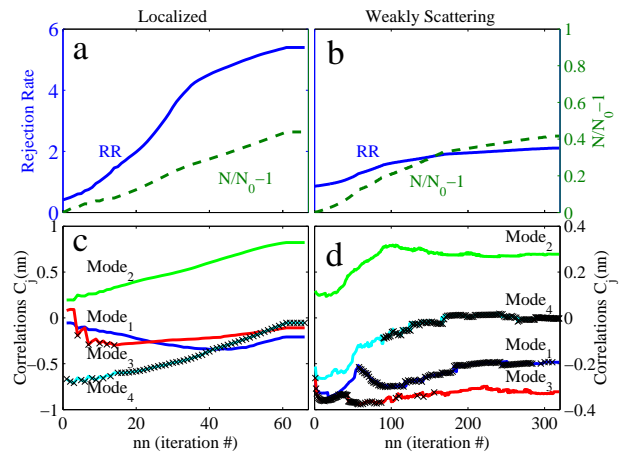}
  \caption{\label{Fig3}
    (Color online) (a,c) localized case, (b,d) weakly scattering case.
    (a,b) Rejection Rate $RR_2$ and threshold $N_2$ for selected mode.
    (c,d) Correlations functions $C_{f{i \in [ 1,3,4 ]}}$.
    Crosses indicates the mode with lowest threshold entering the calculation of $R_2$.
  }
\end{center}
\end{figure}

We then consider $\text{Mode}_2$ ($k = 10.22$ $\mu$m$^{-1}$).
Its rejection rate for uniform pumping ($f_E(x) = 1$) is $RR_2 = 0.45<1$, meaning it would not lase first at threshold.
We now apply the iterative process to select this lasing mode ($\alpha=10$). Its rejection rate increases rapidly as shown in Figure \ref{Fig3}a. It is larger than unity after 10 iterations and converges to $RR_2=5.4$ after 57 iterations. The relative increase of its threshold is less than 50\%.
In contrast, all other modes see their threshold increased by at least an order of magnitude. This is illustrated in Fig.~\ref{Fig2} which shows the evolution of the spectrum in the complex plane. $\text{Mode}_2$ has been efficiently selected. It will be the first to lase above threshold and the singlemode regime will be robust, even at relatively high pumping rate since $RR_2$ is large.
The optimization algorithm has been successfully applied to $\text{Mode}_{i \in [ 1,3,4 ]}$.
Their final rejection rates, $RR_i$ and threshold, $N_i$ are given in \cite{table}.

The optimized pump profile, $f_E(x)$, obtained for $\text{Mode}_{2}$ is shown in Fig.~\ref{Fig2}(d) (see also movie in \cite{table}). It is similar to the lasing mode profile, $g_2(x)$ (Fig.~\ref{Fig2}(c)). The degree of similarity is measured by the spatial correlation $C_{f2}=\langle (f_E-\langle f_E\rangle)(g_2-\langle g_2\rangle) \rangle$, where $f_E$ and $g_2$ have been normalized by their variance, and is close here to unity, $C_{f2}=0.82$. The solution reached by the algorithm is therefore consistent with the predicted efficiency of a local pumping in the localized regime, even in the presence of moderate overlap \cite{VannestePRL01,sebbah02}.
It is also interesting to note that the change of pump profile barely affects the frequency (Fig.~\ref{Fig2}a) and spatial profile (Fig.~\ref{Fig2}c) of the lasing modes, as expected in the localized regime \cite{review}.

We have investigated in more details the working operation of the algorithm by looking at the evolution of the correlations $C_{f{i \in [ 1,2,3,4 ]}}$ of the pump profile with $\text{Mode}_{i \in [ 1,2,3,4 ]}$. As the optimization routine is applied, $C_{f2}$ consistently increases, while correlations $C_{f{i \in [ 1,3,4 ]}}$ of the other modes are progressively minimized, as shown in Fig~\ref{Fig3}c, resulting in increased thresholds for these modes. The crosses in Fig.~\ref{Fig3}c indicate the mode with lowest threshold, $min_{j \ne 2} (N_{j})$, entering the calculation of $RR_2$ at a given iteration. After working alternatively on $\text{Mode}_{3}$ and $\text{Mode}_{4}$, the algorithm works exclusively on the rejection of $\text{Mode}_{4}$. This mode has the largest overlap with $\text{Mode}_{2}$; a fine tuning of the pump profile is therefore required to increase its threshold without increasing too much $\text{Mode}_{2}$'s threshold.

\begin{figure}[ht!]
		        \centering
			 \includegraphics[width=0.8\linewidth]{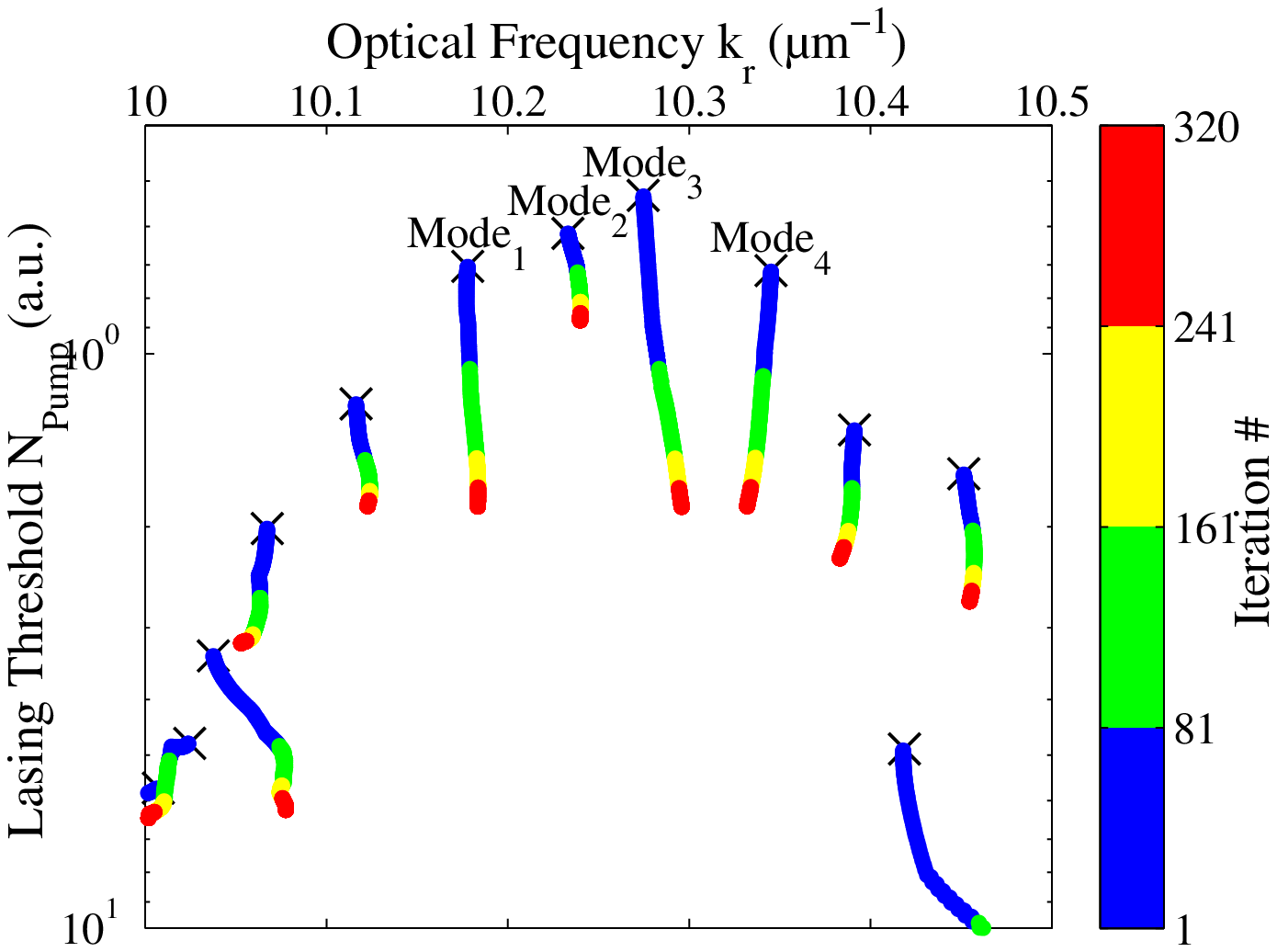}
			\centering
        		\includegraphics[width=0.8\linewidth]{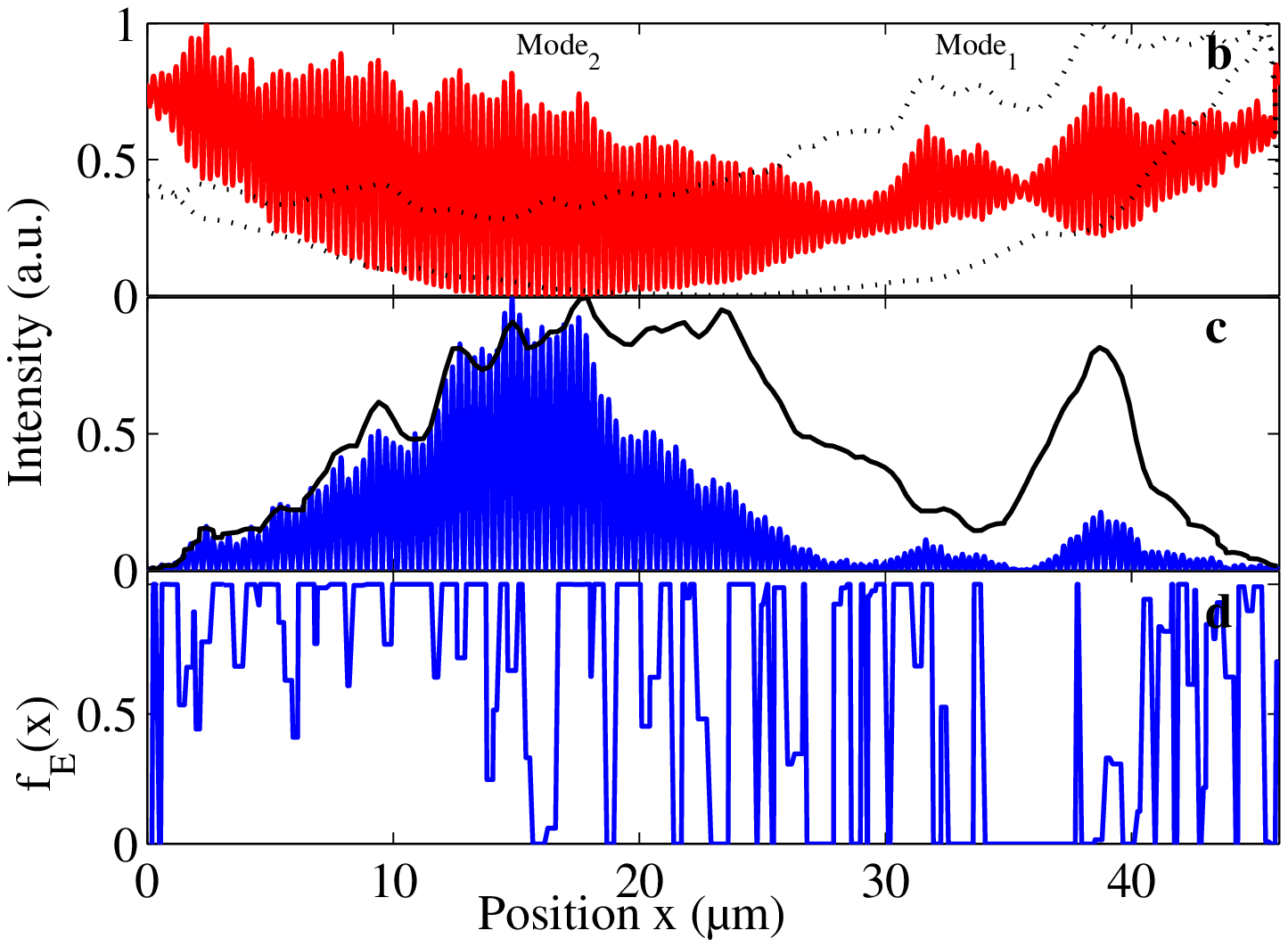}
	\caption{\label{Fig4}
          (Color online) Weakly scattering random laser case.
         (a) Frequency vs. threshold of the lasing modes at successive iterations, when optimization routine is applied to select $\text{Mode}_{2}$.
         Crosses represent initial positions for uniform pumping.
         (b) Spatial profile of $\text{Mode}_1$ (envelope) and $\text{Mode}_2$ (full).
         (c) Stationary component of $\text{Mode}_2$ before (envelop) and after optimization (full).
         (d) Resulting optimized pump spatial profile.
        }
\end{figure}

We turn now to the more difficult case of a weakly scattering random laser.
For an index contrast of $\Delta n = 0.05$, the localization length is $\xi \approx 200 $ $\mu$m $\gg L$
over the frequency range $10.0$ $\mu$m$^{-1} < k < 10.5$ $\mu$m$^{-1}$.
The system is therefore in the weakly scattering regime and lasing modes are spatially extended over the whole system, as illustrated in Fig.~\ref{Fig4}a.

Figures \ref{Fig3}b,d and \ref{Fig4} present the results for the selection of $\text{Mode}_2$ ($k = 10.23$ $\mu m^{-1}$ and ${RR_2}_0=0.86$) ($\alpha=0.3$).
The algorithm converges after 320 iterations with a final rejection rate $RR_2=2.47$.
Although more modest than in the localized case, this increase is significant enough to envision monomode operation of the random laser at that selected frequency,
even when gain saturation is included \cite{andreasenJOSAB11}, since higher threshold modes may be suppressed.
Similar results are obtained for all other modes tested (see \cite{table}).
Figure \ref{Fig4}a shows the relative impact of the iterative process on $\text{Mode}_2$ and on the other modes. As the thresholds increase, lasing modes experience spectral shifts as well as spatial deformation (Fig.~\ref{Fig4}c), in contrast to the localized case. This is a direct consequence of the strong level of pumping required in this case and to the non-uniformity of the gain profile.
Figure \ref{Fig4}d shows the optimized pump profile (see also movie in \cite{table}). Such a profile is unpredictable. A small correlation $C_{f2}=0.28$ is found only when comparing the stationary component of the mode (Fig~\ref{Fig4}c)\cite{andreasennu} and the pump profile. All the value of our adaptative approach is revealed here in this case where modes overlap is significant, threshold distribution is fairly narrow and nonlinear effects strongly impact the modes.
The subtlety of the optimization process is exemplified in Fig~\ref{Fig3}d where the ceaseless switch between modes (crosses) force reduced correlation and even anticorrelation between the rejected modes and the pump profile.

We have intentionally chosen a small index contrast to test the extreme case of a weakly scattering system ($\xi \gg L$).
Intermediate regimes of scattering with various index contrasts have been tested and mode selection by active spatial control worked successfully as well.
Another intentional choice is the optimization criterion because it is weakly constrained. Only two modes are compared at each iteration and as a result, rapid convergence is achieved. Stability studies of the solutions have been performed, which show this optimization method provides most often with local maxima.
Better rejection rates may be obtained, for example, by using a global optimization based on the Ant algorithm \cite{Schluter10}.
We tested it on a a small subset of variables, which yielded promising results.
A full global optimization (over all dielectric layers), however, is numerically out of reach at the moment.
Though the systems studied here with a narrow range of emission frequencies have experimental analogs \cite{bardouxOE11,dardiryAPL11},
typical weakly scattering random lasers yield a huge number of closely packed lasing modes.
In such cases, an efficient global optimization may be necessary, and therefore, is under development.
Finally, it is worthwhile pointing out that the optimization problem posed here is different from \cite{vellekoopPRL08} since the iterative process is highly nonlinear : the medium itself is modified at each iteration by the new computed pump profile. Surprisingly, the alterations of the random laser induced by the pump profile are used to the advantage of the optimization routine, and do not hinder the goal of mode selection.

In summary, we have demonstrated numerically that control of the lasing emission frequency is possible in a random laser.
Active control of the pump profile is proposed to select any lasing mode in the emission spectrum and to significantly increase the threshold of others.
The method has been successfully tested in the regime of strong scattering where modes are spatially localized and can be easily selected using local pumping.
In the weakly scattering regime where strong spatial modal overlap precludes any straightforward spatial selectivity, the algorithm remarkably converges to a complex optimized pump profile which selects the desired lasing mode at the expense of the others.
The proposed algorithm is straightforward to implement in practice.
It also provides with an optimized solution which limits the pump flux on the sample and, for instance, reduce optical damage.
These results open further the way to active control of the lasing properties of random lasers by shaping of the pump profile.
On the basis of our results, we believe output directionality may also be achieved.
For example, local pumping has been shown to yield unidirectional emission in random lasers through the selection of a
new lasing mode \cite{andreasenOL} generated by the presence of gain boundaries \cite{gePRA11}.
Additionally, relaxing the constraint of mode selection may allow a combination of lasing modes to produce the desired directionality.
Moreover, because partial pumping modifies the spatial distributions of lasing modes, the output is not bound by the parameters of the passive random system.
For instance, the frequency shift observed in the weakly diffusive case could be manipulated too to tune at will the emission frequency of the laser.
We believe this approach will foster the interest for random lasers.
But it is not restricted to this field and can be extended e.g. to the domain of high-power broad-area semiconductor lasers where it can be an alternative to the issue of filamentation in order to optimize their brightness \cite{Ohtsubo}.

We thank H. Cao, S. Bhaktha and J. P. Huignard for discussions and suggestions.
This work was supported by the ANR under Grant No. ANR-08-BLAN-0302-01 and the Groupement de Recherche 3219 MesoImage.


\end{document}